\documentclass{ws-procs9x6}

\newcommand{\kin}{\text{kin\,}}
\newcommand{\ext}{\text{ext\,}}
\newcommand{\inter}{\text{int\,}}

\newcommand{\ren}{\text{ren\,}}

\newcommand{\cG}{{\mathcal G}}
\newcommand{\bG}{{\partial\mathcal G}}
\newcommand{\tJ}{{\widetilde{J}}}

\newcommand{\cexG}{\mathcal G_{\text{color}}}
\newcommand{\bJ}{ J_{\partial} }

\newcommand{\cS}{{\mathcal S}}

\newcommand{\bea}{\begin{eqnarray}}
\newcommand{\eea}{\end{eqnarray}}
\newcommand{\beq}{\begin{equation}}
\newcommand{\eeq}{\end{equation}}

\begin{document}

\title{ASYMPTOTIC FREEDOM OF RANK 4 \\ TENSOR GROUP FIELD THEORY\footnote{Contribution to the XXIXth International Colloquium on Group-Theoretical Methods in Physics held at the Chern Institute of Mathematics, Nankai, China, August 20-26, 2012.}
}

\author{J. BEN GELOUN}

\address{Perimeter Institute, 31 Caroline Str N, Waterloo, ON N2L 2Y5,
 Canada\\
E-mail: jbengeloun@perimeterinstitute.ca}

\begin{abstract}
Recently, a rank four tensor group field theory has been 
proved renormalizable. We provide here the key points 
on the renormalizability of this model and 
its UV asymptotic freedom.

\end{abstract}


\bodymatter

\section{Introduction}
\label{aba:sec1}

Random tensor models\cite{Rivasseau:2012yp,Rivasseau:2011xg,Rivasseau:2011hm,oriti,
Gurau:2012vu} have been introduced in the early 90's as a natural 
generalization of matrix models \cite{blaba}. 
Random matrices defined a very successful framework for addressing the quantization of gravity in 2D via random triangulations. 
Until recently, the essential tool used for achieving all analytical results in matrix 
models, namely the t'Hooft 1/N expansion, was crucially missing for
higher rank extension of these models. Indeed, Gurau for a particular class of 
models called colored\cite{color,Gurau:2009tz,Gurau:2010nd} succeeded in defining such a 1/N expansion \cite{Gur3, GurRiv, Gur4,Gurau:2011xp,Gurau:2012vu}.
Important results follow rapidly this breakthrough. To mention a few of these, the statistical analysis of the colored models shows a phase transition with computable critical exponent,\cite{Bonzom:2011zz} longstanding mathematical physics issues \cite{Gurau:2011tj,Gurau:2012ix} 
and universal behavior of these tensors find a sense,\cite{Gurau:2011kk}
and, finally, one defines the first renormalizable tensor models 
 of rank higher than 2.\cite{arXiv:1111.4997,BenGeloun:2012pu,BenGeloun:2012yk,
Geloun:2012bz,Carrozza:2012uv} We refer such a category of quantum field theory to as 
Tensor Group Field Theories (TGFT's).\cite{Rivasseau:2012yp,oriti}

The present work concerns the first of these renormalizable models  addressing the generation
of 4D simplicial pseudo-manifold via a quantum field theoretical formalism.
We provide here an overview of the renormalizability \cite{arXiv:1111.4997} and UV asymptotic freedom\cite{BenGeloun:2012yk} of this model. 
Despite the fact that a contact with a quantization of gravity is not yet understood at this stage,  such a framework is definitely useful for addressing future investigations on emergent geometries.\cite{Rivasseau:2012yp,oriti}

\section{Renormalizability of the model}
\label{sect:renorma}

Consider a complex  rank four tensor over the group
$U(1)$, $\varphi: U(1)^4 \to \mathbb{C}$ which admits the decomposition
in Fourier modes as
\bea
\varphi(h_1,h_2,h_3,h_4)
= \sum_{p_j \in \mathbb{Z}} \varphi_{[p_1,p_2,p_3,p_4]} e^{ip_1 \theta_1} e^{ip_2 \theta_2}
 e^{ip_3 \theta_3} e^{ip_4 \theta_4}
\eea
where   $h_i \in U(1)$, $\theta_i \in [0,2\pi)$ 
and $p_j$ are momentum indices. 
 We denote $\varphi_{[p_1,p_2,p_3,p_4]}=$
$ \varphi_{1,2,3,4}$ and assume no symmetry under permutation of the arguments for this tensor.

The kinetic term of the model, written in momentum space, reads 
\beq
S^{\kin,0} =
 \sum_{p_{j}}
\bar\varphi_{1,2,3,4}
\big[\sum_{s=1}^4 (p_{s})^2 + m^2\big]\varphi_{1,2,3,4} 
\label{skin}
\eeq
where the sum is performed over all momentum values $p_j$.
Clearly, such a kinetic term is inferred from a Laplacian dynamics 
which can be motivated in different ways\cite{Rivasseau:2012yp,Geloun:2011cy}. To this kinetic term, 
we associate a Gaussian measure $d\mu_C[\varphi]$ with covariance $C = (\sum_s p^2_s + m^2)^{-1}$.

Trace invariant objects \cite{Gurau:2012ix}
define the interaction part of the model:
\bea 
S_{6;1} &=& \sum_{p_{j}}
\varphi_{1,2,3,4} \bar\varphi_{1',2,3,4}\varphi_{1',2',3',4'} 
 \bar\varphi_{1'',2',3',4'} \varphi_{1'',2'',3'',4''} \bar\varphi_{1,2'',3'',4''} 
+ \text{perm.} 
\cr\cr
S_{6;2} &=& \sum_{p_{j}}
\varphi_{1,2,3,4} \bar\varphi_{1',2',3',4}\varphi_{1',2',3',4'} 
\bar\varphi_{1'',2,3,4'}\varphi_{1'',2'',3'',4''} \bar\varphi_{1,2'',3'',4''}
+ \text{perm.}\crcr
&& \label{interac6} \\
  S_{4;1} & =&  \sum_{p_{j}} \varphi_{1,2,3,4} \,\bar\varphi_{1',2,3,4}\,\varphi_{1',2',3',4'} \,\bar\varphi_{1,2',3',4'}
+ \text{perm.}  
\label{S41}\\
S_{4;2}&=&
\big[\sum_{p_j} \bar\varphi_{1,2,3,4} \,
\varphi_{1,2,3,4} \big]
\big[\sum_{p'_j} \bar\varphi_{1',2',3',4'}\, 
\varphi_{1',2',3',4'} \big]   
\label{fifi}
\eea
where the sum (+perm.) is over all 24 permutations of the four color indices. The flow of $S_{6;2}$ generates a $\phi^4$-type interaction that 
we will refer to as  {\it anomalous term} \eqref{fifi} and so should be part of our action.

\begin{figure}[ht]
\begin{center}
 \includegraphics[width=3cm, height=0.8cm]{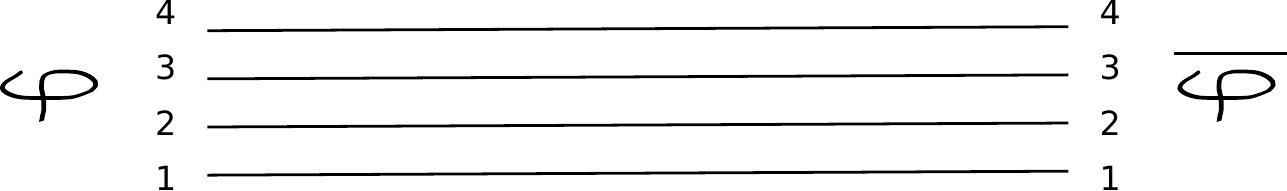}  
\caption{{\small The propagator. }}
\label{fig:prop}
\end{center}
\end{figure}

\begin{figure}
 \centering
     \begin{minipage}[t]{.8\textwidth}
      \centering
\includegraphics[angle=0, width=7cm, height=2.5cm]{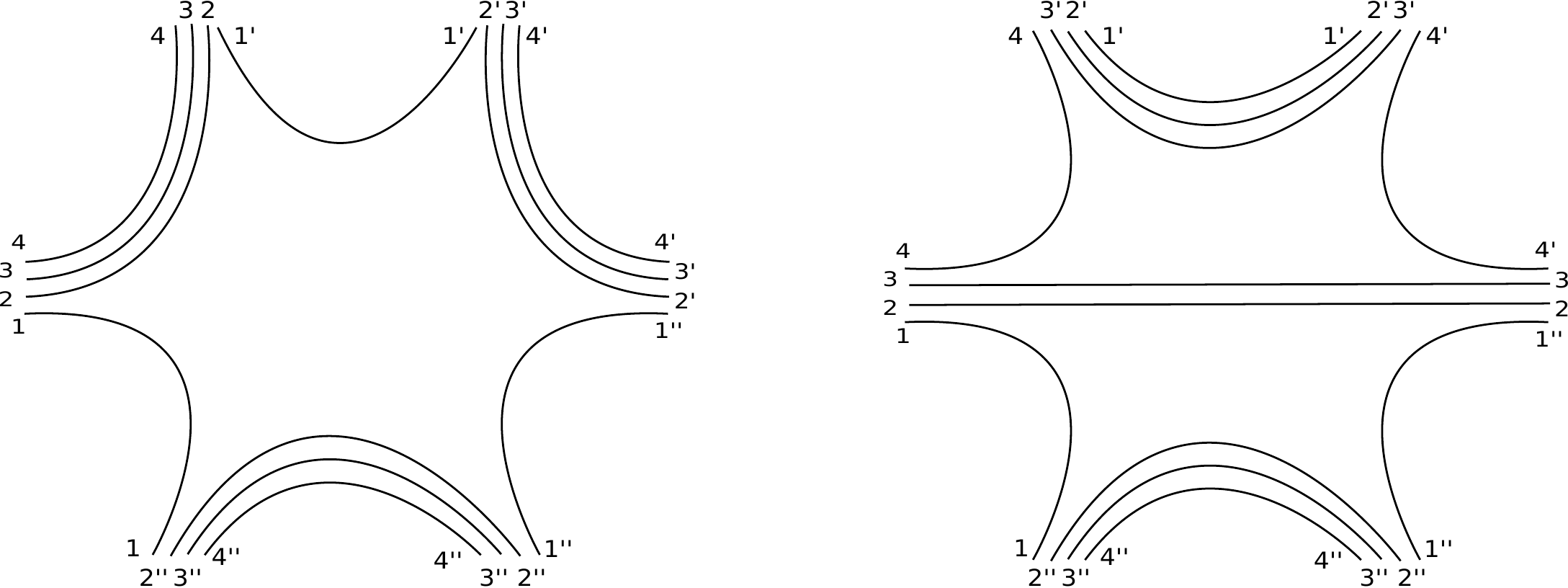}
\caption{ {\small Vertices of the  type $\phi^6_{(1)}$ from $S_{6;1}$ (left)  and of the type $\phi^6_{(2)}$ from  $S_{6;1}$ (right). \label{vert6}}}
\end{minipage}
\end{figure}

\begin{figure}
 \centering
     \begin{minipage}[t]{.8\textwidth}
      \centering
\includegraphics[angle=0, width=7cm, height=2.5cm]{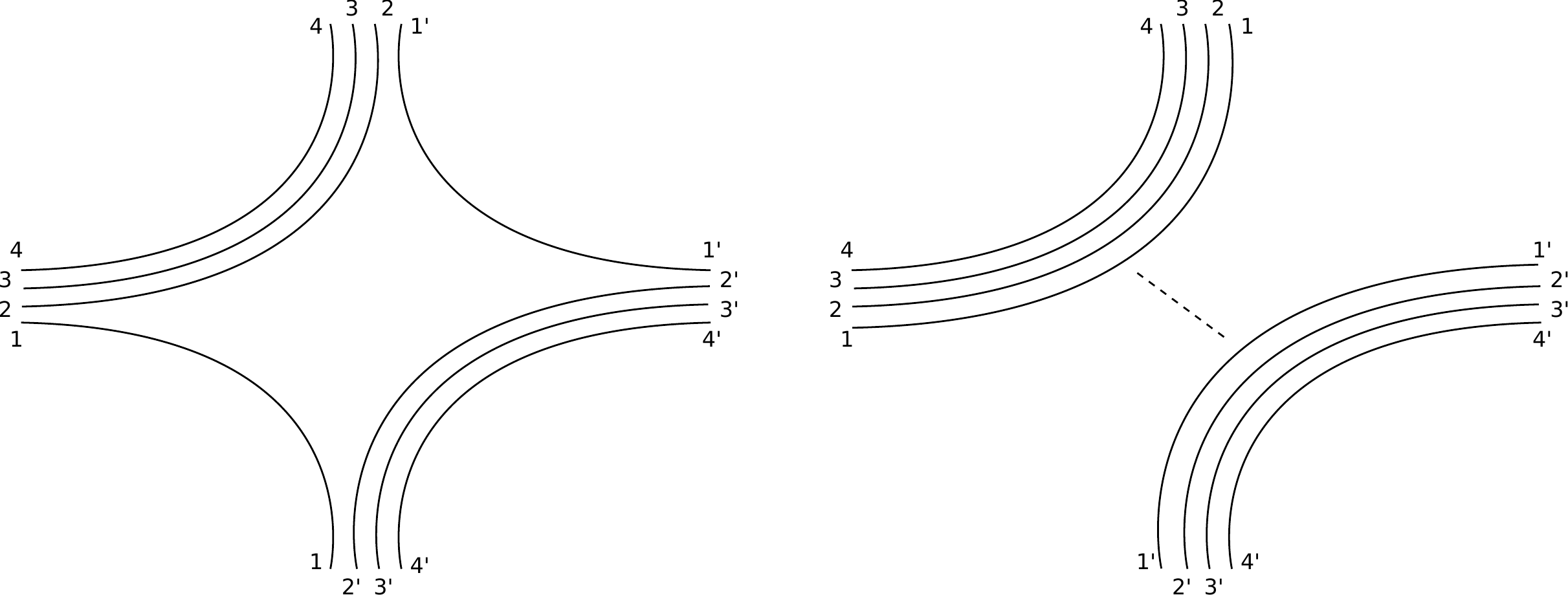}\\
\caption{ {\small Vertices of the type $\phi^4_{(1)}$ from $S_{4;1}$ (left) and the anomalous term $\phi^4_{(2)} = (\phi^2)^2$ from $S_{4;2}$ (right).
\label{vert4}}}
\end{minipage}
\end{figure}

Feynman graphs have a tensor structure. Lines are four-stranded and represent propagators  (Figure \ref{fig:prop}) 
whereas vertices are nonlocal objects (as depicted in Figure \ref{vert6} and Figure  \ref{vert4}). 

Introducing an ultraviolet cutoff $\Lambda$ and counterterms, one proves\cite{arXiv:1111.4997} that the model described
by
\eqref{skin} and \eqref{interac6}-\eqref{fifi} is renormalizable 
at all orders of perturbation theory. This statement is proved by  multiscale analysis \cite{Rivasseau:1991ub} and the study of the graph topology. 
We review now the main ingredients of this proof. 

$\bullet$ {\emph{A slice decomposition}}: Use the Schwinger representation of the propagator $C=\int_{0}^\infty e^{-\alpha (\sum_s p^2_s + m^2)} d\alpha$, to slice
this quantity as $C=\sum_{i=0}^{\infty} C_i$ where $C_i=\int_{M^{-2i}}^{M^{-2i+2}} 
e^{-\alpha (\sum_s p^2_s + m^2)} d\alpha\leq K M^{-2i}e^{M^{-2i}(\sum_{s}p^{2}+m^2)}$, for $M>0$.

$\bullet$ {\emph{The multi-scale expansion}} of a graph amplitude using the above bound yields the {\emph{crude power-counting}} at a momentum attribution $\mu$, $A_{\cG;\mu}\leq  K^n \prod_{(i,k)} M^{\omega_d(G^i_k)}$, where $K,n$ are some constants, $G^i_k$ are called the quasi-local subgraphs\cite{Rivasseau:1991ub} of $\cG$, $\omega_d(G) = -2L(G)+F_{\inter}(G)$ is the degree of divergence of the graph $G$, $L(G)$ the number of internal lines and $F_{\inter}(G)$  the number of closed loops or faces of the subgraph $G$. Quasi-local subgraphs
define the {\emph{generalized notion of locality}} here. 

$\bullet$ {\emph{A refined power-counting theorem}}:
 A graph $\cG$ in this theory admits
a five-color extension $\cexG$ which is itself a rank four tensor graph.
A jacket $J$ of $\cexG$ is a ribbon subgraph of $\cexG$ defined by a color cycle $(0abcd)$ up to a cyclic permutation. 
 The jacket $\tJ$ is the closed jacket obtained from 
$J$ after closing all external legs present in $J$. 
The boundary $\bG$ of the graph $\cG$
is the closed graph defined by vertices corresponding to 
external legs and by lines corresponding to external strands of $\cG$ 
\cite{Gurau:2009tz}. It is, presently, a vacuum rank $3$ colored graph.
 A boundary jacket $\bJ$ is a jacket of  $\bG$. 

The anomalous vertex $(\phi^2)^2$ is
disconnected from the point of view of its strands. 
We define {\em connected} component graphs by reducing such vertices
to twice $\phi^2$-vertices.

The divergence degree any connected graph $\cG$ is given by
\beq
\label{contopformula}
\omega_d(\cG) = -{\textstyle{\frac13}} \big[ \sum_{J} g_{\tJ} 
-  \sum_{\bJ} g_{\bJ} \big] - [C_{\bG}-1] 
- V_4- 2[V_2 +V_2''] - {\textstyle{\frac12}} \left[ N_{\ext}- 6\right]
\eeq
where $g_{\tJ} $ and $g_{\bJ}$ are the genus of $\tJ$ 
and $\bJ$, respectively, $C_{\bG}$ is the number
of connected components of the boundary graph $\bG$; the first sum is performed on all closed jackets 
$\tJ$ of $\cexG$  and the second sum is performed on
all boundary jackets $\bJ$ of $\bG$;
$V_4$ is the number of $\phi^4_{(1)}$ vertices,
$V_2$ the number of vertices of  the type $\phi^2$ (mass counterterms),
$V''_2 = 2V_4' $ is twice the number of vertices of 
type $\phi^4_{(2)}=(\phi^2)^2$, $N_{\ext}$ its number
of external legs. 

 One notices that $\omega_d(\cG)$ does not
depend on vertices of the type $\phi^6$. Finally, the following table lists all primitively divergent graphs:

\begin{center} {\scriptsize{
\begin{tabular}{lccccc||cc|}
$N_{\ext}$ & $V_2 + V_2''$ & $V_4$ & $\sum_{\bJ} g_{\bJ}$ & $C_{\bG}-1$ & $\sum_{\tJ}  g_{\tJ}$ & $\omega_d(\cG)$  \\
\hline\hline
6 &0 & 0 & 0 & 0& 0& 0\\
\hline
4 & 0 & 0 & 0 & 0 & 0 & 1 \\
4 & 0  & 1 & 0 & 0 & 0 & 0\\
4 & 0 & 0 & 0 & 1 & 0 & 0 \\
\hline
2 & 0 & 0 & 0 & 0 & 0 & 2\\
2 & 0 & 1 & 0 & 0 & 0 & 1\\
2 & 0 & 2 & 0 & 0 & 0 & 0\\
2 & 0 & 0 & 0 & 0 & 6 & 0\\
2 & 1 & 0 & 0 & 0 & 0 & 0\\
\hline\hline
\end{tabular} 

\vspace{0.2cm}
Table 1: List of primitively divergent graphs }}
\end{center}

Call  graphs satisfying
$\sum_{\tJ} g_{\tJ}=0$ ``melonic'' graphs or simply ``melons'' \cite{Bonzom:2011zz}. Thus, some graphs in Table 1 are melons with melonic boundary.

\section{$\beta$-functions}
\label{sect:beta}

We will discuss the $\beta$-functions of the model given by the interaction
\bea
S = \textstyle{\frac13}\lambda_{6;1}\,S_{6;1} + \lambda_{6;2} \, S_{6;2} +
 \textstyle{\frac12} \lambda_{4;1} \, S_{4;1} 
+\textstyle{\frac12}\lambda_{4;2}\, S_{4;2} 
 \eea
where some symmetry factors are introduced. 
It turns out that the behavior of the renormalized
coupling  constants $\lambda^\ren_{6;1/2}$ in the UV 
are those significant for this model. 

The $\beta$-functions of the $\phi^6_{(1/2)}$ are evaluated 
with two ingredients: (1) the truncated and amputated one particle irreducible (1PI) 6-point functions $\Gamma_{6;1/2}(\cdot)$ the external data of which are designed in the form of the initial (bare) interaction;
(2) the wave function renormalization given by 
$Z = 1 - \partial_{b_1^2}\Sigma|_{b_{1,2,3,4} = 0}$, 
where
$\Sigma(b_1,b_2,b_3,b_4) = \langle \varphi_{1,2,3,4}
\bar\varphi_{1,2,3,4}\rangle^t_{1PI}$  is the so-called self-energy 
or sum of all amputated 1PI two-point functions. 
We identify all relevant graphs related to the six- and two-point 
functions using Table 1.

The following ratios encode the $\beta$-functions of the coupling constants:
\beq
\label{gamma6}
\lambda_{6;1/2}^{\ren} = -
\textstyle{\frac{\Gamma_{6;1/2}(0,0,0,0,0,0,0,0,0,0,0,0)}{ Z^{3}} }
\eeq
At two loops (for the first) and four loops (for the second), one finds that  the renormalized  coupling constants satisfy the equations
\bea
&& 
\lambda^\ren_{6;2} = \lambda_{6;2} 
+ 2\lambda_{6;2} \lambda_{6;1}S^1 
 +3 \lambda_{6;2}^2   [S^1 +  S^{12}] +O(\lambda^3)
\label{lam12} \\
&&
 \lambda^\ren_{6;1} = \lambda_{6;1}  + 
8 \lambda_{6;1}^3\cS    + O(\lambda_{6;1}^4)
\eea
where $S^{1,12}$ and $\cS$ are formal log-divergent sums.\cite{BenGeloun:2012pu} The $\beta$-functions
of this model at this order of perturbation are given by 
\bea
\beta_{6;2;\; (2)} = 3 \quad \qquad 
\beta_{6;2;\; (12)} = 2 \quad\qquad
\beta_{6;1} = 8
\label{beta}
\eea
Assuming positive coupling constants, \eqref{lam12}-\eqref{beta}
show that the overall model is asymptotically free in UV.

In fact, the complete study of all renormalized coupling equations shows that\cite{BenGeloun:2012pu} there exists a UV 
fixed manifold for the model given by
\beq
\lambda_{6;1/2} = 0\qquad  \forall \lambda_{4;1} \qquad
\lambda_{4;2} = 0
\eeq
Perturbing the system around this fixed manifold by adding
small quantities to the bare couplings, the renormalized coupling constants increase in the infrared (IR). This fact generally hints at a phase transition, a scenario very promising for finding new models in the  continuum.

\section*{Acknowledgements}
 The organizers of the XXIXth International Colloquium on Group-Theoretical Methods in Physics, Nankai, China, are warmly thank for their welcome and hospitality. Discussions with  R. Gurau and V. Rivasseau are gratefully acknowledged. 
Research at Perimeter Institute is supported by the Government of Canada through Industry 
Canada and by the Province of Ontario through the Ministry of Research and Innovation.


\begin{thebibliography}{00}



\bibitem{Rivasseau:2012yp} 
  V.~Rivasseau,
  ``The Tensor Track: an Update,''
(a contribution to the same proceeding)
  arXiv:1209.5284 [hep-th].

\bibitem{Rivasseau:2011xg}
  V.~Rivasseau,
  ``Towards Renormalizing Group Field Theory,''
  PoS C {\bf NCFG2010}, 004 (2010)
  [arXiv:1103.1900 [gr-qc]].


\bibitem{Rivasseau:2011hm} 
  V.~Rivasseau,
  ``Quantum Gravity and Renormalization: The Tensor Track,''
  arXiv:1112.5104 [hep-th].

 \bibitem{oriti}
  D.~Oriti,
  ``The group field theory approach to quantum gravity,''
  arXiv:gr-qc/0607032.

\bibitem{Gurau:2012vu} 
  R.~Gurau,
  ``A review of the large N limit of tensor models,''
(a contribution to the same proceeding)
  arXiv:1209.4295 [math-ph].

  R.~Gurau,
  ``A review of the 1/N expansion in random tensor models,''
  arXiv:1209.3252 [math-ph].


\bibitem{blaba}
  P.~Di Francesco, P.~H.~Ginsparg and J.~Zinn-Justin,
 ``2-D Gravity and random matrices,''
  Phys.\ Rept.\  {\bf 254}, 1 (1995)
  [arXiv:hep-th/9306153].

\bibitem{color}
 R.~Gurau,
  ``Colored Group Field Theory,''
  Commun.\ Math.\ Phys.\  {\bf 304}, 69 (2011)
  [arXiv:0907.2582 [hep-th]].

\bibitem{Gurau:2009tz}
  R.~Gurau,
  ``Topological Graph Polynomials in Colored Group Field Theory,''
  Annales Henri Poincare {\bf 11}, 565 (2010)
  [arXiv:0911.1945 [hep-th]].

\bibitem{Gurau:2010nd}
  R.~Gurau,
  ``Lost in Translation: Topological Singularities in Group Field Theory,''
  Class.\ Quant.\ Grav.\  {\bf 27}, 235023 (2010)
  [arXiv:1006.0714 [hep-th]].

\bibitem{Gur3}
  R.~Gurau,
 ``The 1/N expansion of colored tensor models,''
  Annales Henri Poincare {\bf 12}, 829-847 (2011).
 [arXiv:1011.2726 [gr-qc]].

\bibitem{GurRiv}
  R.~Gurau and V.~Rivasseau,
 ``The 1/N expansion of colored tensor models in arbitrary dimension,''
  Europhys.\ Lett.\  {\bf 95}, 50004 (2011).
  [arXiv:1101.4182 [gr-qc]].


\bibitem{Gur4}
  R.~Gurau,
  ``The complete 1/N expansion of colored tensor models in arbitrary dimension,''
  Annales Henri Poincare {\bf 13}, 399 (2012)
  [arXiv:1102.5759 [gr-qc]].

\bibitem{Gurau:2011xp} 
  R.~Gurau and J.~P.~Ryan,
  ``Colored Tensor Models - a review,''
  SIGMA {\bf 8}, 020 (2012)
  [arXiv:1109.4812 [hep-th]].

\bibitem{Bonzom:2011zz}
  V.~Bonzom, R.~Gurau, A.~Riello and V.~Rivasseau,
  ``Critical behavior of colored tensor models in the large N limit,''
  Nucl.\ Phys.\  B {\bf 853}, 174 (2011)
  [arXiv:1105.3122 [hep-th]].

\bibitem{Gurau:2011tj}
  R.~Gurau,
  ``A generalization of the Virasoro algebra to arbitrary dimensions,''
  Nucl.\ Phys.\  B {\bf 852}, 592 (2011)
  [arXiv:1105.6072 [hep-th]].



\bibitem{Gurau:2012ix} 
  R.~Gurau,
  ``The Schwinger Dyson equations and the algebra of constraints of random tensor models at all orders,''
  arXiv:1203.4965 [hep-th].


\bibitem{Gurau:2011kk}
  R.~Gurau,
  ``Universality for Random Tensors,''
  arXiv:1111.0519 [math.PR].

\bibitem{arXiv:1111.4997} 
  J.~Ben Geloun and V.~Rivasseau,
  ``A Renormalizable 4-Dimensional Tensor Field Theory,''
  arXiv:1111.4997 [hep-th]; Commun. Math. Phys. 2012
DOI 10.1007/s00220-012-1549-1; 

  J.~Ben~Geloun and V.~Rivasseau,
``Addendum to 'A Renormalizable 4-Dimensional Tensor Field Theory',''
  arXiv:1209.4606 [hep-th]; to appear Commun. Math. Phys. 2012.


\bibitem{BenGeloun:2012pu} 
  J.~Ben Geloun and D.~O.~Samary,
  ``3D Tensor Field Theory: Renormalization and One-loop $\beta$-functions,''
  arXiv:1201.0176 [hep-th].


\bibitem{BenGeloun:2012yk} 
  J.~Ben Geloun,
``Two and four-loop $\beta$-functions of rank 4 renormalizable tensor field theories,''
  arXiv:1205.5513 [hep-th].


\bibitem{Geloun:2012bz} 
  J.~Ben~Geloun and E.~R.~Livine,
  ``Some classes of renormalizable tensor models,''
  arXiv:1207.0416 [hep-th].



\bibitem{Carrozza:2012uv} 
  S.~Carrozza, D.~Oriti and V.~Rivasseau,
  ``Renormalization of Tensorial Group Field Theories: Abelian U(1) Models in Four Dimensions,''
  arXiv:1207.6734 [hep-th].

\bibitem{Rivasseau:1991ub}
  V.~Rivasseau, 
  ``From perturbative to constructive renormalization,''
Princeton series in physics (Princeton Univ. Pr., Princeton, 1991).


\bibitem{Geloun:2011cy}
  J.~Ben Geloun and V.~Bonzom,
  ``Radiative corrections in the Boulatov-Ooguri tensor model: The 2-point function,''
  Int.\ J.\ Theor.\ Phys.\  {\bf 50}, 2819 (2011)
  [arXiv:1101.4294 [hep-th]].


\end{thebibliography}
\end{document}